\documentclass[aps,prx,reprint,superscriptaddress,longbibliography,footinbib]{revtex4-1} 
\usepackage{graphicx}
\usepackage{amsmath,amssymb}
\usepackage{textcomp,siunitx}
\DeclareSIUnit\gauss{G}
\usepackage[
	pdftex,
	colorlinks=true,
  	urlcolor=blue,
  	linkcolor=blue,
  	citecolor=blue
]{hyperref}
\usepackage{color}

\newcommand{\ket}[1]{\left\lvert #1 \right\rangle}%
\newcommand{\bra}[1]{\left\langle #1 \right\lvert}%
\newcommand{\avg}[1]{\left\langle #1 \right\rangle}%
\newcommand{\abs}[1]{\left\lvert #1 \right\rvert}%

\begin{document}

\title{Quench Dynamics of a Fermi Gas with Strong Non-Local Interactions}
\author{Elmer Guardado-Sanchez}
\affiliation{Department of Physics, Princeton University, Princeton, New Jersey 08544 USA}
\author{Benjamin M. Spar}
\affiliation{Department of Physics, Princeton University, Princeton, New Jersey  08544 USA}
\author{Peter Schauss}
\affiliation{Department of Physics, University of Virginia, Charlottesville, Virginia 22904 USA}
\author{Ron Belyansky}
\affiliation{Joint Quantum Institute, NIST/University of Maryland, College Park, Maryland 20742, USA}
\affiliation{Joint Center for Quantum Information and Computer Science, NIST/University of Maryland, College Park, Maryland 20742 USA}
\author{Jeremy T. Young}
\affiliation{Joint Quantum Institute, NIST/University of Maryland, College Park, Maryland 20742, USA}
\affiliation{JILA, NIST and Department of Physics, University of Colorado, Boulder, Colorado 80309, USA}
\affiliation{Center for Theory of Quantum Matter, University of Colorado, Boulder, Colorado 80309, USA}
\author{Przemyslaw Bienias}
\affiliation{Joint Quantum Institute, NIST/University of Maryland, College Park, Maryland 20742, USA}
\affiliation{Joint Center for Quantum Information and Computer Science, NIST/University of Maryland, College Park, Maryland 20742 USA}
\author{Alexey V. Gorshkov}
\affiliation{Joint Quantum Institute, NIST/University of Maryland, College Park, Maryland 20742, USA}
\affiliation{Joint Center for Quantum Information and Computer Science, NIST/University of Maryland, College Park, Maryland 20742 USA}
\author{Thomas Iadecola}
\affiliation{Department of Physics and Astronomy, Iowa State University, Ames, Iowa 50011, USA}
\author{Waseem S. Bakr}
\email[Corresponding author. Email: ]{wbakr@princeton.edu}
\affiliation{Department of Physics, Princeton University, Princeton, New Jersey  08544 USA}
\date{\today}

\begin{abstract}
We induce strong non-local interactions in a 2D Fermi gas in an optical lattice using Rydberg dressing. The system is approximately described by a $t-V$ model on a square lattice where the fermions experience isotropic nearest-neighbor interactions and are free to hop only along one direction. We measure the interactions using many-body Ramsey interferometry and study the lifetime of the gas in the presence of tunneling, finding that tunneling does not reduce the lifetime. To probe the interplay of non-local interactions with tunneling, we investigate the short-time relaxation dynamics of charge density waves in the gas. We find that strong nearest-neighbor interactions slow down the relaxation. Our work opens the door for quantum simulations of systems with strong non-local interactions such as extended Fermi-Hubbard models.
\end{abstract}

\maketitle

\section{INTRODUCTION}

Ultracold gases are a versatile platform for studying quantum many-body physics \cite{Bloch2008}. The ability to engineer and control the interactions in these systems has played an important role in observing novel phases of matter including crossover fermionic superfluids \cite{Inguscio2008} and dipolar supersolids \cite{Bottcher-Pfau2019,Tanzi-Mudugno2019,Chomaz-Ferlaino2019} and in studying out-of-equilibrium dynamical processes such as thermalization \cite{Tang-Lev2018}. Recent efforts have focused on degenerate quantum gases with long-range interactions including those of magnetic atoms~\cite{Bottcher-Pfau2019,Tang-Lev2018,Patscheider-Mark2020,Gabardos-Vernac2020} and polar molecules \cite{DeMarco-Ye2019,Bohn2017}. These systems may be distinguished from other quantum platforms with long-range interactions including ions \cite{Blatt-Roos2012,Bruzewicz2019}, Rydberg atoms \cite{Saffman2010}, polar molecules in optical tweezers \cite{Anderegg-Doyle2019,Liu-Ni2019} and atoms in optical cavities \cite{Esslinger2013}, in that the particles are itinerant. This can lead to an interesting interplay between interactions, kinetic energy and quantum statistics. Rydberg dressing has been proposed as an alternative route to realize quantum gases with tunable long-range interactions \cite{Pupillo-Zoller2010, Johnson-Rolston2010, Henkel-Pohl2010}. Experimental demonstrations of Rydberg dressing \cite{Balewski-Pfau2014,Goldschmidt-Porto2016, Aman-Burgdorfer2016, Jau-Biedermann2016, Zeiher-Gross2016, Gaul-Pohl2016, Boulier-Porto2017, Zeiher-Gross2017, Arias-Whitlock2019, Borish-Schleier-Smith2020} have been performed with localized atoms or quantum gases of heavy atoms where observation of motional effects has been elusive. 

However, the combination of motion and {Rydberg dressing} can lead to novel phenomena and shed new light on {the many-body physics of} spinless and spinful fermionic systems with {power-law} interactions.
In 1D, Rydberg dressing leads to quantum liquids with qualitatively new features relative to the Tomonaga-Luttinger liquid paradigm~\cite{Mattioli-Pupillo2013}.
In 2D, topological Mott insulators can be emulated by placing atoms on a Lieb lattice~\cite{Dauphin-Martin-Delgado2016}.
Compared to contact or on-site interactions, the long-range interactions between Rydberg-dressed atoms makes it easier to achieve the low filling factors required for quantum Hall states~\cite{Grass2018,Burrello2020}.
The interplay between hole motion and antiferromagnetism---believed to be at the heart of high-temperature superconductivity---can be studied in Rydberg-dressed atomic lattices emulating the $t-J_z$ model~\cite{Grusdt2018a}.
In 3D, one can achieve exotic topological density waves~\cite{Li2015b}, topological superfluids~\cite{Xiong2014}, and metallic quantum solid phases~\cite{Li2016d}.

Here we investigate Rydberg dressing of lithium-6, a light fermionic atom. Its fast tunneling in an optical lattice allows us to study the quench dynamics of itinerant fermions with strong, purely off-site interactions.

Atoms in a quantum gas resonantly coupled to a Rydberg state experience strong van der Waals interactions many orders of magnitude larger than their kinetic energy for typical interatomic spacings, hindering access to the interesting regime where the two energy scales compete. At the same time, the population of atoms in the Rydberg state decays on a timescale of tens of microseconds, short compared to millisecond motional timescales. Rydberg dressing addresses both of these issues. Using an off-resonant coupling, the atoms are prepared in a laser-dressed eigenstate $\ket{g_{\textrm{dr}}}\approx \ket{g} + \beta\ket{r}$ of predominant ground state ($\ket{g}$) character and a small Rydberg ($\ket{r}$) admixture, where $\beta=\frac{\Omega}{2\Delta}\ll1$, $\Omega$ is the coupling strength, and $\Delta$ is the laser detuning from the transition frequency. This enhances the lifetime of the dressed atom by a factor of $1/\beta^2$ relative to the bare Rydberg state lifetime. On the other hand, the interaction between two atoms a distance $r$ apart is reduced in strength and can be approximately described by a tunable softcore potential $V(r)=V_{\textrm{max}}/(r^6+r_c^6)$ with strength $V_{\textrm{max}}\sim \beta^3 \Omega$ and range $r_c\sim (|C_6/2\Delta|)^{1/6}$, where $C_6$ is the van der Waals coefficient for the Rydberg-Rydberg interaction. Early experiments with 3D quantum gases were limited by rapid collective atom loss attributed to a blackbody-induced avalanche dephasing effect \cite{Balewski-Pfau2014, Goldschmidt-Porto2016, Aman-Burgdorfer2016, Boulier-Porto2017}. Nevertheless, Rydberg dressing has been successfully used to entangle atoms in optical tweezers \cite{Jau-Biedermann2016}, perform electrometry in bulk gases \cite{Arias-Whitlock2019}, and study spin dynamics \cite{Zeiher-Gross2016, Zeiher-Gross2017, Borish-Schleier-Smith2020}.

In this work, we report on the single-photon Rydberg dressing of a 2D $^6$Li Fermi gas in an optical lattice in the presence of tunneling. This results in a lattice gas of fermions with strong, non-local interactions. We characterize the interaction potential using many-body Ramsey interferometry \cite{Zeiher-Gross2016}. A careful study of the lifetime of spin-polarized gases shows different behavior compared to previous Rydberg dressing realizations, with the lifetime depending strongly on the density but not on the atom number at fixed density. We also observe that the presence of tunneling in the system has no effect on the lifetime. Finally, we use this platform to realize a 2D coupled-chain $t-V$ model consisting of interaction-coupled chains and study the short-time quench dynamics of charge-density wave states, finding that the strong attractive interactions inhibit the motion of the atoms.

Theoretical studies of the 1D $t-V$ model~\cite{Tomasi-Pollmann2019,Yang2019d} have shown that it can exhibit Hilbert-space fragmentation (HSF)~\cite{Sala20,Khemani-Nandkishore2020}, in which dynamical constraints ``shatter" the Hilbert space into exponentially many disconnected subspaces. Like many-body localization (MBL)~\cite{Nandkishore15,Abanin19} and quantum many-body scars~\cite{Bernien-Lukin2017,Turner-Papic2018}, HSF is a mechanism whereby isolated quantum systems can fail to reach thermal equilibrium after a quantum quench~\cite{DAlessio-Rigol2016}. In the $t-V$ model, HSF arises in the limit of strong interactions, where the number of ``bond" excitations, i.e., nearest-neighbor pairs of fermions, joins the total fermion number as a conserved quantity. Our mixed-dimensional $t-V$ model inherits properties of the 1D version, including the HSF in the limit $t/V\rightarrow 0$. Our quench results demonstrate experimentally how HSF impacts the short-time relaxation dynamics for nonzero $t/V$.


\begin{figure}
\includegraphics[width=\columnwidth]{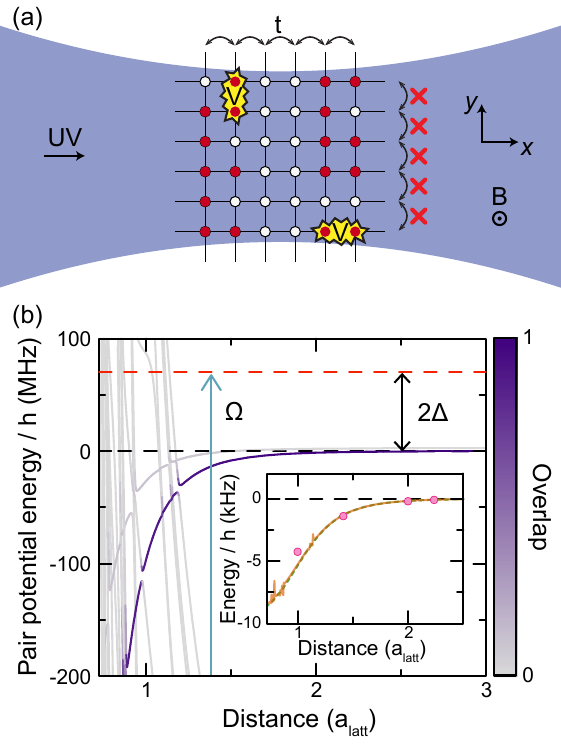}
\caption{{\bfseries Realization of a $t-V$ model with Rydberg dressing.} (a) The Rydberg dressing beam propagates along the $x$-direction of the lattice, effectively decoupling 1D chains in the $y$-direction due to a differential light shift. Hopping of fermions (red dots) along the $x$-direction is unaffected. Interactions are isotropic. (b) $^6$Li pair potentials for dressing to the state $\ket{28P, m_l=0, m_s=-1/2}$ calculated using \cite{Weber2017}. The color of the lines represents the overlap with the target pair-state ($\ket{28P,0,-1/2}\otimes\ket{28P,0,-1/2}$) coupled via the laser with Rabi coupling $\Omega$ and detuning $\Delta$ from the target state. Inset: Calculated dressed potential for $\Omega = 2\pi\times \SI{7.66}{\MHz}$ and $\Delta = 2\pi\times\SI{35}{\MHz}$ taking into account the overlaps to all pair potentials (orange solid line). The dashed green line represents the expected dressed potential for a simple van der Waals potential with $C_6 = h\times\SI{90.19}{\MHz~a_{latt}^6}$. Pink points are the interaction at each lattice distance taking into account the wavefunction spread of the atoms. \label{fig:fig1}}
\end{figure}

\section{Experimental system and theoretical model}

Our system consists of a degenerate Fermi gas of $^6$Li atoms in a square optical lattice of spacing $a_{\textrm{latt}} = \SI{752}{\nm}$ (Fig.~\hyperref[fig:fig1]{~\ref*{fig:fig1}a})~\cite{Guardado-Bakr2018}. We apply a $\SI{591.8(3)}{G}$~\footnote{$\SI{1}{Gauss}=\SI{1e-4}{Tesla}$} magnetic field perpendicular to the 2D system. We load spin-polarized gases prepared in a state that may be labeled at high fields as $\ket{nl,m_l,m_s,m_I}=\ket{2S,0,-1/2,1}=\ket{1}$, or alternatively $\ket{2S,0,-1/2,-1}=\ket{3}$ depending on the measurement. We have control over the initial density profile by employing a spatial light modulator. Using a $\SI{231}{\nm}$ laser beam with linear polarization parallel to the magnetic field and propagating along the lattice $x$-direction, we couple the ground state atoms to the $\ket{28P,0,-1/2}$ Rydberg state (App.~\ref{app:expdetails}). By tuning the intensity and the detuning of the dressing light~\footnote{The laser is locked to an ultralow expansion glass cavity and we can set the detuning with an uncertainty of $\sim2\pi\times\SI{200}{\kHz}$ by referencing to resonant blowing of a sparse system at very low intensity which is limited by the linewidth of the laser.}, we have real-time control over the isotropic soft-core interaction potential between the atoms in the gas (Fig.~\hyperref[fig:fig1]{\ref*{fig:fig1}b}).

The lattice system is described by a single-band spinless fermion Hamiltonian
\begin{equation}
	\hat{H} = -t \sum_{\langle i,j\rangle} (\hat{c}_i^\dagger \hat{c}_j + \textrm{h.c.}) + \sum_{i\neq j} \frac{V_{ij}}{2} \hat{n}_i\hat{n}_j + \sum_i \delta_{i} \hat{n}_i,
	\label{eq:Hamiltonian}
\end{equation}
where $t$ is a tunneling matrix element, $V_{ij}$ is the off-site interaction [Eq.~\ref{eq:eqB3} and Fig.~\hyperref[fig:fig1]{\ref*{fig:fig1}b(inset)}] and $\delta_{i}$ is the potential due to single-particle light shifts contributed by the lattice and Rydberg dressing beams. Since our dressing beam is tightly focused with a waist of $\SI{16.1(4)}{\mu m}$, the change in $\delta$ between rows in the $y$-direction, which is orthogonal to the beam propagation axis, is much larger than $t$ (for typical experiments presented in Sec.~\ref{sec:dynamics}, the minimum change in $\delta$ between rows is $> 3t$ near the intensity maximum of the Rydberg dressing beam). On the other hand, because of the large Rayleigh range of the beam ($\sim\SI{3.5}{mm}$), the variation of $\delta$ along the beam propagation direction ($x$-direction) is negligible. To first approximation, we drop the light shift term and the hopping along the $y$-direction. Thus, we can rewrite our Hamiltonian as a coupled-chain $t-V$ model of the form
\begin{align}
\label{eq:tV-Ham}
	\hat{H} = -t \sum_{\langle i,j\rangle_x} (\hat{c}_i^\dagger \hat{c}_j + \textrm{h.c.}) + \sum_{i\neq j} \frac{V_{ij}}{2} \hat{n}_i\hat{n}_j.
\end{align}


\section{Characterization of the system}
\subsection{Rydberg-dressed interaction potentials}

\begin{figure}
\includegraphics[width=\columnwidth]{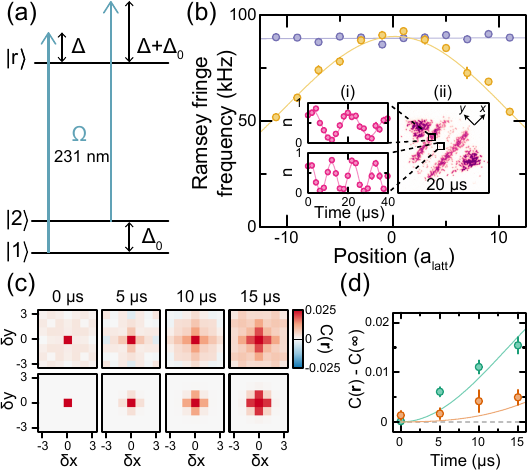}
\caption{{\bfseries Measuring Rydberg dressed interactions with many-body Ramsey interferometry.} (a) Energy level diagram for $^6$Li showing that the dressing of the other hyperfine ground state cannot be ignored. Here $\Delta/2\pi$ is varied between \SI{30}{\MHz} and \SI{100}{\MHz} while $\Delta_0/2\pi=\SI{75.806(1)}{\MHz}$. (b) Ramsey fringe frequency measured at a detuning of $\Delta = 2\pi\times\SI{100}{\MHz}$ at different positions in the cloud. The frequency is almost constant along the propagation direction of the beam (purple). In the transverse direction (yellow), it varies rapidly as expected for a tightly focused Gaussian beam. Insets: (i) Ramsey oscillations at two representative positions in the cloud. (ii) Sample image of one spin state in the cloud at $T = \SI{20}{\mu s}$. (c) Spin correlations for different spin-echo pulse times at $\Omega = 2\pi\times\SI{7.66(7)}{\MHz}$ and $\Delta = 2\pi\times\SI{35}{\MHz}$. Measurement (top) and theoretical expectation (bottom). (d) Nearest (green) and next-nearest (orange) neighbor correlations after subtracting $C(\infty)$. Lines correspond to the expected correlations. Experimental error bars correspond to standard error of the mean. \label{fig:fig2}}
\end{figure}

In order to characterize the Rydberg dressing interaction potentials, we perform many-body Ramsey interferometry between states $\ket{1}$ and $\ket{2}=\ket{2S,0,-1/2,0}$ following the procedure introduced in Ref. \cite{Zeiher-Gross2016}. Starting from a spin-polarized band insulator of atoms prepared in state $\ket{1}$ in a deep lattice that suppresses tunneling, a $\pi/2$ radiofrequency pulse prepares a superposition of state $\ket{1}$ and $\ket{2}$, which acquire a differential phase during a subsequent evolution for time $T$ in the presence of the dressing light. Unlike Ref. \cite{Zeiher-Gross2016}, the splitting between the hyperfine ground-states of $^6$Li is comparable to the detuning $\Delta$ of the dressing laser (Fig.~\hyperref[fig:fig2]{\ref*{fig:fig2}a}), and both states are significantly dressed by the light (App.~\ref{app:2dressing}). First, we obtain the spatial profile of the Rabi coupling strength $\Omega(i,j)$ by measuring the population of $\ket{2}$ after a $\pi/2-T-\pi/2$ pulse sequence using a detuning $\Delta = 2\pi\times\SI{100}{\MHz}$. The large detuning is chosen so that the interactions, whose strength scales as $1/\Delta^3$, are negligible, while the single-particle light shifts that scale as $1/\Delta$ lead to a large differential phase during the evolution. From these measurements, we extract the waist of the beam ($\SI{16.1(4)}{\mu m}$) and measure Rabi couplings up to $\Omega = 2\pi \times \SI{9.48(8)}{\MHz}$ (Fig.~\hyperref[fig:fig2]{\ref*{fig:fig2}b}). The measured spatial profile of the Ramsey fringe frequency confirms the rapid variation of $\delta_i$ along the $y$-direction, while no variation of $\delta_i$ is observed along the $x$-direction within the statistical uncertainty of the measurement ($\sim 1$ kHz).

To probe interactions in the system, we switch to a smaller detuning $\Delta = 2\pi\times\SI{35}{\MHz}$. We measure density correlations of state $\ket{1}$ ($C(\textbf{r}) = \avg{n_1(\textbf{r})n_1(0)}-\avg{n_1(\textbf{r})}\avg{n_1(0)}$) after a spin-echo pulse sequence ($\pi/2-T-\pi-T-\pi/2$) which eliminates differential phases due to the light shift. Fig.~\hyperref[fig:fig2]{\ref*{fig:fig2}c} shows the measured correlations after different evolution times $T$ compared to the theoretical expectation (App.~\ref{app:ramsey}).  Fig.~\hyperref[fig:fig2]{\ref*{fig:fig2}d} depicts the evolution of the nearest-neighbor and next-nearest-neighbor correlations with the correlation offset $C(\infty)$ subtracted. This offset is attributed to correlated atom number fluctuations in the images \cite{Zeiher-Gross2016}. We find good agreement with the theoretical model, which predicts a nearest-neighbor (next-nearest-neighbor) attractive interaction $|V_{10}|=h\times$4.2(2) kHz ($|V_{11}|=h\times$1.37(6) kHz)  (Fig.~\hyperref[fig:fig1]{\ref*{fig:fig1}b})

\subsection{Lifetime}
\label{sec:lifetime}

To probe coherent many-body physics in our system, the lifetime $\tau$ of the sample has to be larger than the interaction and tunneling times. Atoms resonantly excited to a Rydberg state are lost from our system on a timescale of tens of microseconds for several reasons: photon recoils due to spontaneous emission and large forces due to anti-trapping optical potentials and due to interactions with other Rydberg atoms. Due to its Rydberg admixture, an isolated dressed atom decays with a lifetime  $\tau_{\textrm{eff}} = \tau_0/\beta^2$, where $\tau_0$ is the lifetime of the Rydberg state determined by radiative and blackbody-driven transitions to other states. Previous experiments with frozen 2D and 3D systems have observed much shorter lifetimes than $\tau_{\textrm{eff}}$ \cite{Balewski-Pfau2014,Goldschmidt-Porto2016, Aman-Burgdorfer2016, Boulier-Porto2017,Zeiher-Gross2016}. A simplified model used to explain these experiments considers a blackbody-driven decay of the dressed state to a pure Rydberg state of opposite parity. The first such contaminant appears in the system on a timescale $\tau_c=\tau_{\textrm{BB}}/(N\beta^2)$ where $\tau_{\textrm{BB}}$ is the blackbody lifetime of the Rydberg state and $N$ is the number of atoms in the system. This atom interacts with other dressed atoms through resonant state-exchange characterized by a $C_3$ coefficient, broadening the Rydberg line. In particular, other atoms at a certain facilitation radius $(|C_3/\Delta|)^{1/3}$ will be resonantly excited, leading to avalanche loss of all the atoms from the trap. Experiments in 2D have indeed observed a collective lifetime close to $\tau_c$ and a bimodal atom number distribution in lifetime measurements \cite{Zeiher-Gross2016}. We have not observed such bimodality in our 2D systems, and the lifetime does not depend strongly on $N$ at fixed density (App.~\ref{app:lifetimeVnumber}). In this regard, our 2D $^6$Li experiments are closer to $^{87}$Rb experiments with 1D chains where the avalanche mechanism is suppressed to some extent \cite{Zeiher-Gross2017}.

\begin{figure}
\includegraphics[width=\columnwidth]{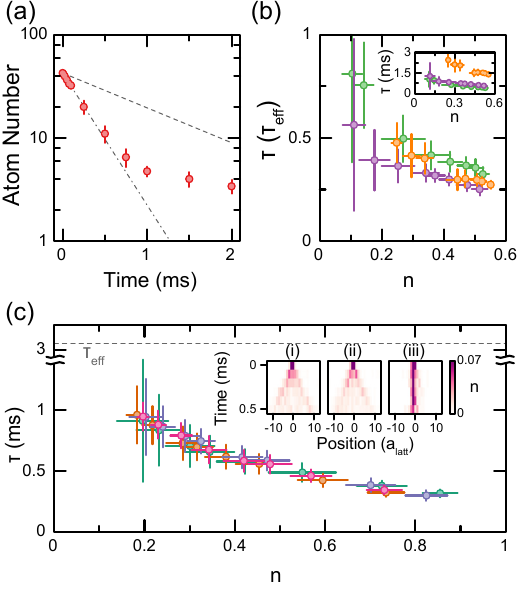}
\caption{{\bfseries Lifetime of itinerant Rydberg dressed fermions.} (a) Atom number vs. dressing time for a frozen gas. The red circles correspond to measurements on a system of 7 by 7 sites. Dashed-dotted line corresponds to an exponential fit to the first 5 data points and dashed line corresponds to the expected single-particle dressed lifetime $\tau_{\textrm{eff}}$. (b) Measured lifetime in a frozen gas in units of $\tau_\textrm{eff}$ vs. the initial density for $\Omega=2\pi\times\SI{9.25(8)}{\MHz}$ and $\Delta=2\pi\times\left(30~\mathrm{(green)}, 40~\mathrm{(purple)}, 60~\mathrm{(orange)}\right)\SI{}{\MHz}$. Inset: Same measurements in units of $\SI{}{ms}$. (c) Lifetime vs. initial density for different tunnelings: $\SI{0.01}{\kHz}$ (green), $\SI{0.25}{\kHz}$ (purple), $\SI{1.0}{\kHz}$ (orange), and $\SI{1.7}{\kHz}$ (pink). The data is taken with $\Omega=2\pi \times \SI{6.04(8)}{\MHz}, \Delta=2\pi\times \SI{30}{\MHz}$. Insets: (i) Tunneling dynamics of atoms sparsely initialized on a strip along the $y$-direction with no dressing light. From this data, we extract a tunneling rate $t = h\times\SI{1.7}{\kHz}$. (ii) Same measurement in the presence of the dressing light. (iii) Same measurement in the presence of the dressing light but with the strip along the $x$-direction. Experimental error bars correspond to standard error of the mean.  \label{fig:fig3}}
\end{figure}

The atom number decay in a frozen system of 7 by 7 sites is shown in Fig.~\hyperref[fig:fig3]{\ref*{fig:fig3}a}. The decay is not exponential, indicating a density-dependent lifetime which we extract by fitting different sections of the decay curve. For dressing to $\ket{28P}$, $\tau_0=\SI{30.5}{\us}$ \cite{Beterov-Entin2009}. We measured the density-dependent lifetime for $\Omega=2\pi \times 9.25(8)$ MHz at three different detunings, $\Delta=2\pi \times\left(30, 40, 60\right)$ MHz (Fig.~\hyperref[fig:fig3]{\ref*{fig:fig3}b}). Around half-filling, the collective lifetime is $\sim0.3\tau_{\textrm{eff}}$ for $\Delta=2\pi \times 30$ MHz and approaches $\tau_{\textrm{eff}}$ for the smallest densities ($n\sim 0.1$). For comparison, perfect avalanche loss would predict $\tau_c=0.08\tau_{\textrm{eff}}$. 

Next, we measure the lifetime of the dressed gas in the presence of tunneling, which has been a topic of theoretical debate \cite{Li-Lesanowsky2013, Macri-Pohl2014}. We measure the density-dependent lifetime for different lattice depths, spanning the frozen gas regime to a tunneling of 1.7 kHz (Fig.~\hyperref[fig:fig3]{\ref*{fig:fig3}c}). We do not observe any change of the lifetime with tunneling. A potential concern in this measurement is that the tunneling along the $x$-direction may be suppressed by uncontrolled disorder in $\delta_i$. We rule this out by preparing a sparse strip of atoms and observing its tunneling dynamics. As expected for a clean dressed system, the tunneling dynamics along the $x$-direction is almost identical to the case without the dressing light, while the dynamics is frozen along the $y$-direction (Fig.~\hyperref[fig:fig3]{\ref*{fig:fig3}c} inset). Combining the results of our interferometry and lifetime measurements, we achieve a lifetime of several interaction times measured by the figure of merit $ V_{\textrm{10}}\tau/\hbar \sim 20$~\cite{Zeiher-Gross2017} for a mobile system with $n=0.5$.

\section{Quench Dynamics}
\label{sec:dynamics}

\begin{figure}
\includegraphics[width=\columnwidth]{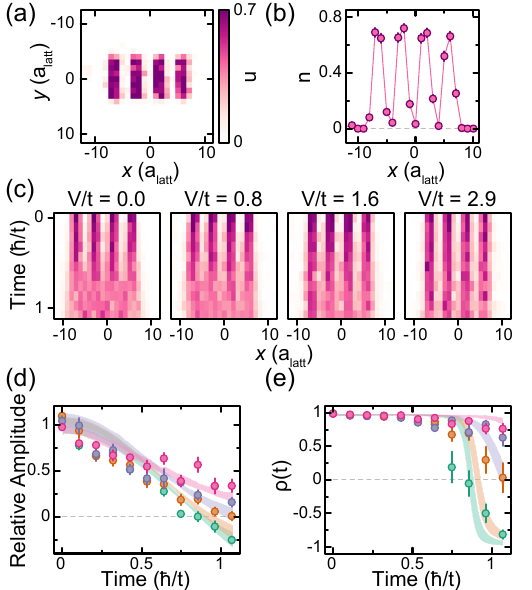}
\caption{{\bfseries Interaction dependence of quench dynamics of a charge density wave.} (a) Average initial state density profile for the quench measurements. (b) Density profile averaged along the $y$-direction of the initial state shown in (a). (c) Density profile time evolution for interactions $V/t = [0, 0.78(7), 1.61(8), 2.9(2)]$. Color scale is the same as in (a). (d) Fitted relative amplitude of density profile vs. time. Colors (green, orange, purple, and pink) correspond to the interactions in (c) from lowest to highest. (e) Autocorrelation function of the density pattern. Colors are same as in (d). Shaded curves correspond to numerical simulations. Experimental error bars correspond to standard error of the mean.   \label{fig:fig4}}
\end{figure}

To probe the interplay of interactions and tunneling in our system, we use light patterned with a spatial light modulator to initialize the system in a charge density wave state of atoms in state $\ket{3}$. The initial density pattern approximates a square wave with period $\lambda=\SI{4}{a_{latt}}$ and width $w=\SI{7}{a_{latt}}$, with the average density oscillating between $n\sim0$ and $n\sim0.7$. (see Figs.~\hyperref[fig:fig4]{\ref*{fig:fig4}a-b}). Dynamics in a lattice with $t=h\times\SI{1.7}{\kHz}$ is initiated by suddenly turning off the patterning potential while keeping walls in the $y$-direction as in \cite{Guardado-Bakr2020}. We average the density profiles over the non-hopping direction and observe a qualitative change in the dynamics as we increase $V/t$ (here $V\equiv |V_{10}|$) from 0 to 2.9(2) (Fig.~\hyperref[fig:fig4]{\ref*{fig:fig4}c}). To emphasize the evolution of the pattern, we scale the data to account for atom loss during the evolution (App.~\ref{app:quenchlifetime}). In the non-interacting quench, we observe that the phase of the charge density wave inverts at a time $\sim \hbar/t$ as is expected for a coherent evolution \cite{Brown-Bakr2019}. For strong interactions, the decay of the charge density wave slows down and the system retains a memory of its initial state for longer times.

This can be understood as an interplay between two conservation laws: the intrinsic U(1) particle number ($\hat{N}=\sum_x\hat{n}_x$) conservation as well as an emergent conservation of the number of bonds $\hat{N}_b=\sum_{x}\hat{n}_x\hat{n}_{x+1}$. The latter becomes a conserved quantity  when the longer range interactions are ignored, and in the limit of infinite $V/t$. States of the form $\ket{...0011001100...}$ along the hopping direction, which the imprinted density pattern attempts to approximate, would be completely frozen in the limit of infinite $V/t$ \cite{Tomasi-Pollmann2019}. For a large but finite $V$, moving a single atom (and hence breaking a bond) costs an energy of up to $3V$, which is energetically unfavorable, and hence leads to reduced relaxation dynamics. 

To quantify the difference in the dynamics of the different quenches, we employ two different methods. The first is to fit a sinusoid of the form $n(x,t) = A\sin{(2\pi x/\lambda + \phi)} + B$ to determine the amplitude of the wave relative to its mean, $A/B$ (Fig.~\hyperref[fig:fig4]{\ref*{fig:fig4}d}). The fit is restricted to $\abs{x}\leq \SI{6}{a_{latt}}$, and $\phi$ is fixed by the initial pattern. The second method is to calculate the autocorrelation function
\begin{align}
    \rho(t) &= \frac{\mathrm{cov}_x(n(x,0),n(x,t))}{\mathrm{\sigma}_x(n(x,0))\mathrm{\sigma}_x(n(x,t))},
\end{align}
where cov$_x$ and $\sigma_x$ are the covariance and the standard deviation respectively (Fig.~\hyperref[fig:fig4]{\ref*{fig:fig4}e}).

\begin{figure}
\includegraphics[width=\columnwidth]{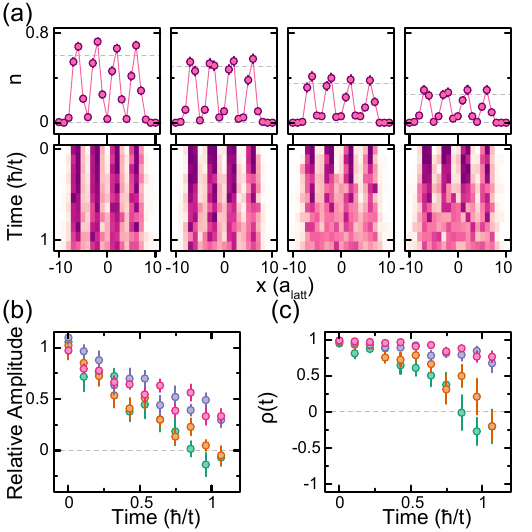}
\caption{{\bfseries Density dependence of quench dynamics.} (a) (top) Initial state density profiles. (bottom) Corresponding time evolution of each initial state for $V/t=2.9(2)$. Colorbar is same as Fig.~\hyperref[fig:fig4]{\ref*{fig:fig4}a} with limits set by dotted lines on top panel. (b) Fitted relative amplitude of density profiles vs. time. Colors (green, orange, purple, and pink) correspond to the initial states in (a) from low to high density. (c) Autocorrelation function of the density pattern. Colors are same as in (b). Experimental error bars correspond to standard error of the mean. \label{fig:fig5}}
\end{figure}

Further confirmation that the slower decay of the charge density waves is an interaction effect is obtained by varying the average density in the initial state. Fig.~\hyperref[fig:fig5]{\ref*{fig:fig5}} shows these initial states and their time evolution for $V/t=2.9(2)$. As the average density of the initial state is decreased, it approaches a ``sparse'' limit where the probability of having two neighboring atoms is negligible. In this regime, the system is effectively non-interacting and we recover the phase inversion during the evolution. Since these measurements are done at fixed power of the dressing light, they rule out disorder-induced localization as a mechanism for arresting the dynamics.


\section{Numerical simulations}

We use exact diagonalization to simulate the quench dynamics of our experiment. As the simulation for the full experimental 2D system ($\sim 7\times21$) is computationally intractable, we compare instead to numerics on a $2\times11$ $t-V$ model with only nearest-neighbor interactions and no tunneling along the $y$-direction and find qualitative agreement with the measurements.

We account for atom loss during the experiment via a Lindblad master equation $\partial_t \hat{\rho} = -i(\hat{H}_{\textrm{eff}}\hat{\rho}-\hat{\rho}\hat{H}_{\textrm{eff}}^\dagger)+\Gamma\sum_i \hat{a}_i\hat{\rho}\hat{a}^\dagger_i$. Here, $\hat{H}_{\textrm{eff}}=\hat{H}-i\frac{\Gamma}{2}\hat{N}$ is the effective non-Hermitian Hamiltonian [$\hat{H}$ is the $t-V$ Hamiltonian from Eq.~(\ref{eq:tV-Ham})] and the second term describes quantum jumps corresponding to atom loss with rate $\Gamma$.
We solve the master equation using the quantum trajectory approach \cite{DaleyTrajs}. Note that the anti-Hermitian term in $\hat{H}_{\textrm{eff}}$ is a constant due to the particle number conservation, and hence it can be neglected since $\hat{H}_{\textrm{eff}}$ and $\hat{H}$ generate the same dynamics (up to the normalization, which only serves to determine the timings of the quantum jumps).

The initial state for each trajectory is sampled directly from the experimental data taken at $t=0$. We pick a $2\times 9$ region centered on 2 of the 4 density peaks from the experimental images (Fig.~\ref{fig:fig4}a). In order to reduce boundary effects, we add empty sites on each end of the chain. We average the resulting dynamics over the different trajectories, whose number is comparable to the number of experimental snapshots. Next, we analyze the averaged simulated dynamics using the same methods we use for the experimental data. Fig.~\ref{fig:fig4} shows the comparison of the experiments with these numerical simulations. We find good qualitative agreement with this small 2D coupled-chain numerical model.

The 2D nature of the system is important for fully understanding the relaxation time-scales in our system. In particular, in a one-dimensional system, moving a single atom from the initial ``...00110011..." pattern (and hence breaking a bond) costs an energy $V$. However, in the coupled-chain $t-V$ model with isotropic interaction, breaking a bond now costs up to $3V$ for the idealized initial charge density wave state. We thus expect the 2D system to have slower relaxation rate compared to a 1D system with the same interaction strength.

\begin{figure}
\includegraphics[width=\columnwidth]{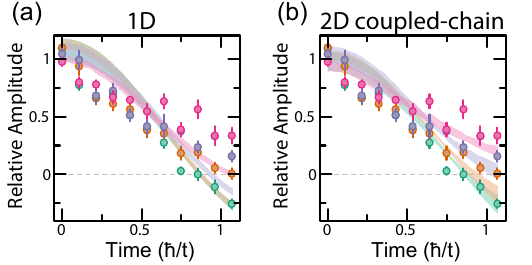}
\caption{{\bfseries Role of interchain couplings in slowing down charge density wave relaxation.} Numerical simulations of a $t-V$ model with tunneling $t$ along only one direction and isotropic nearest-neighbor interactions $V$. (a) Fitted relative sinusoid amplitude to observed (circles) and calculated quench dynamics of $1\times21$ systems (shaded regions). The colors represent the different interaction strengths $V/t = [0~\mathrm{ (green)}, 0.78(7)~\mathrm{ (orange)}, 1.61(8)~\mathrm{ (purple)}, 2.9(2)~\mathrm{ (pink)}]$ explored in the experiment. (b) Same comparison as in (a) but for calculations done on $2\times11$ systems. This is Fig.~4d. Experimental error bars correspond to standard error of the mean. \label{fig:sfig4}}
\end{figure}

To verify this, we perform additional numerical simulations on a single chain of 21 atoms. Similarly to our 2D simulation, we sample $1\times19$ arrays from the experimental snapshots at $t=0$ and add empty sites at the ends. We find that the atoms spread quicker than they do in the ladder geometry and have worse agreement with the experimental results. Fig.~\ref{fig:sfig4} shows a comparison between the 1D and 2D coupled-chain numerical simulations on the one hand and the experimental data on the other. This comparison highlights the importance of the interchain interactions in order to fully understand our system.

The remaining discrepancy between some of the numerical and experimental results could be attributed to several factors. First, we are only able to simulate a smaller system than in the experiment. We expect that adding additional chains could further slow down the relaxation dynamics. Second, our modelling of the atom loss via a Lindblad master equation assumes that the decay rate is exponential. However, as we showed in Sec.~\ref{sec:lifetime}, the decay rate is actually non-uniform in space and time, and depends on the density.

\section{Conclusions}

Our results present a new frontier in quantum simulations of itinerant lattice models with strong off-site interactions. By working with larger $r_c/a$, spinless fermion models may be used to explore equilibrium phases such as topological Mott states \cite{Dauphin-Martin-Delgado2012} or cluster Luttinger liquid phases \cite{Mattioli-Pupillo2013}. Moreover, the system considered in this work provides a platform for the experimental realization of models prevalent in theoretical studies of non-equilibrium dynamics. For example, the 1D $t-V$ model can be mapped to the XXZ spin chain, which has long been studied in the context of many-body localization~\cite{Znidaric08,Bardarson12,Serbyn13}. This model and variants thereof have also been proposed to harbor dynamical phases intermediate between full MBL and thermalization~\cite{Luitz15,Schecter18,Rakovszky-Pollmann2020}. Our work lays the foundation for future studies of such phenomena, as well as other non-equilibrium dynamical regimes including prethermalization~\cite{Nessi-Cazalilla2014}. Furthermore, the close spacing between the hyperfine ground states of $^6$Li also opens the door for the simultaneous dressing of two spin states and the exploration of extended Fermi-Hubbard models.

The present experiment has allowed us to start probing coherent dynamics in $t-V$ models, which we plan to continue to explore especially upon improving the interaction-lifetime figure of merit. For example, for small but finite $t/|V|$, it is possible to access a complex hierarchy of timescales for quench dynamics that depends crucially on the initial state~\cite{Yang2019d}.

Our work motivates further theoretical and experimental exploration of the mixed-dimensional models in the context of both the non-equilibrium dynamics and ground-state physics~\cite{Grusdt2018} such as meson formation. Another promising direction based on the interplay of Rydberg-dressing and atomic motion is vibrational dressing~\cite{Mazza2020,Gambetta2019a}, non-destructive cooling~\cite{Belyansky2019a},  an exploration of multi-band physics, as well as the use of microwave-dressed Rydberg states, allowing for both attractive and repulsive dressed $1/r^3$ dipole-dipole interactions \cite{Young2020}.

There are several possible approaches to improve the interaction-lifetime figure of merit. Enhancement of the Rabi coupling by over an order of magnitude may be achieved using a build-up cavity \cite{Cooper-Yost2018}. For a single-particle system, the figure of merit scales with $\Omega$ at fixed $\beta$, while further enhancement of the collective lifetime is expected in this regime due to shrinking facilitation radii for increasing $\Delta$. Increasing $\Omega$ by a factor of 10 at fixed $\beta$ leads to facilitation radii that are a factor of $10^{1/3}$ smaller. For almost all states coupled to by blackbody radiation, the facilitation radii become less than one site. If collective loss is completely inhibited, the combined effect is to enhance the figure of merit by a factor of $\sim30$. The principal quantum number used in this experiment was chosen to keep the range of the interaction on the order of one site. Relaxing this constraint or alternatively using a larger lattice spacing would allow using longer-lived Rydberg states at higher principal quantum number. Using electric fields to tune close to a F\"{o}rster resonance results in deep potential wells that may be exploited to enhance the figure of merit by a factor of $|\Delta|/\Omega$ \cite{Binjen-Pohl2015} and potentially allow us to achieve repulsive interactions. Finally, the single particle lifetime can be improved and the collective black-body induced atom loss may be completely eliminated by operating at cryogenic temperatures (improving the figure of merit by a factor of $\sim6$ for fixed dressing laser parameters).

\begin{acknowledgments}
We thank David Huse, Alan Morningstar, Zhi-Cheng Yang, Seth Whitsitt, and Fangli Liu for helpful discussions. We also thank Zoe Yan, Adam Kaufman and Ana Maria Rey for feedback on the manuscript. This work was supported by the NSF (grant no. DMR-1607277), the David and Lucile Packard Foundation (grant no. 2016-65128), and the AFOSR Young Investigator Research Program (grant no. FA9550-16-1-0269). W.S.B. was supported by an Alfred P. Sloan Foundation fellowship. R.B., J.T.Y., P.B., and A.V.G. acknowledge funding by AFOSR, AFOSR MURI, DoE ASCR Quantum Testbed Pathfinder program (award No. DE-SC0019040), U.S. Department of Energy Award No. DE-SC0019449, DoE ASCR Accelerated Research in Quantum Computing program (award No. DE-SC0020312), NSF PFCQC program, ARL CDQI, ARO MURI, and NSF PFC at JQI. R.B. was also supported by fellowships from the NSERC and FRQNT of Canada. J.T.Y was supported in part by the NIST NRC Research Postdoctoral Associateship Award. Specific product citations are for the purpose of clarification only, and are not an endorsement by NIST.
\end{acknowledgments}


\appendix{}

\section{Experimental Details}\label{app:expdetails}
The experimental setup and basic parameters are described in detail in the supplement of Ref.~\cite{Brown-Bakr2017}. The procedure for calibrating the spatial light modulator is described in the supplement of Ref~\cite{Brown-Bakr2019}. The $\SI{231}{nm}$ UV laser system for Rydberg excitation is described in the supplement of Ref.~\cite{Guardado-Bakr2018}. 

\subsection*{Power stabilization of the Rydberg dressing light}
The UV light used for Rydberg dressing is generated using a $\SI{923}{nm}$ amplified diode laser system followed by two second harmonic generation cavities in series. The fractional power stability of the UV light after the second cavity is about $10 \%$ which was sufficient for our previous work with direct excitation to Rydberg states \cite{Guardado-Bakr2018}. However, in the case of dressing, power stability is more critical due to the interaction strength having a quartic dependence on the Rabi frequency ($V\propto \Omega^4$). Furthermore, the stability of the power during the spin-echo sequence used in the Ramsey interferometry is important to cancel the phases accumulated due to the single-particle light shift. We improved the power stability to much better than 1 \% by adding a noise-eater. The noise-eater consists of an electro-optic polarization modulator (QuBig PCx2B-UV) and an $\alpha$-BBO Glan-Taylor polarizer (EKSMA 441-2108). By measuring the laser power using a pick-off before the last acousto-optic modulator (\cite{Guardado-Bakr2018}) and feeding back on the noise-eater, we suppressed intensity noise for frequencies up to $\SI{1}{\MHz}$ and eliminated shot-to-shot drifts in the dressing light intensity that limited our previous experiments.

\begin{figure}
\includegraphics[width=\columnwidth]{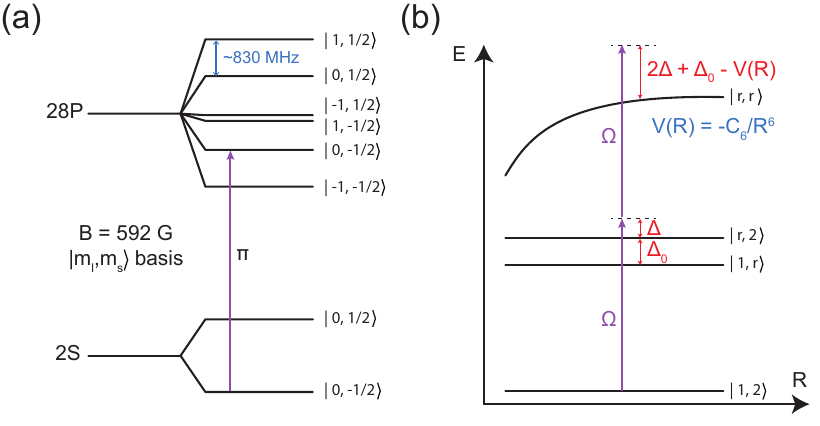}
\caption{{\bfseries  Rydberg dressing of $^6$Li.} (a) Level diagram showing the hyperfine ground states of $^6$Li directly coupled to the $28P$ Rydberg state using linearly ($\pi$) polarized light at a field of $\SI{592}{G}$. The basis used is $\ket{m_l,m_s}$. (b) Rydberg dressing scheme for two atoms in different hyperfine ground states $\ket{1}$ and $\ket{2}$ coupled to the Rydberg state $\ket{r}$. $\Omega$ is the Rabi coupling of the laser, $\Delta$ is the detuning from the resonant transition between $\ket{1}$ and $\ket{r}$, $\Delta_0$ is the hyperfine splitting between $\ket{1}$ and $\ket{2}$ and $V(R)=-C_6/R^6$ is the van der Waals interaction potential between two Rydberg states $\ket{r}$.  \label{fig:sfig1}}
\end{figure}

\subsection*{Ground and Rydberg states used in the experiments}
We work at a magnetic field of $\SI{592}{G}$ pointing in the direction perpendicular to the 2D lattice plane. At this field, both the ground and Rydberg states are in the Paschen-Back regime such that we can approximately label them using the $\ket{nl,m_l,m_s,m_I}$ basis (Fig.~\hyperref[fig:sfig1]{\ref*{fig:sfig1}a}). As explained in the text, the hyperfine ground states we use are $\ket{1}, \ket{2}$ and $\ket{3}$ numbered from lowest to highest in energy and having $m_I=1,0,-1$ respectively. For the Rydberg states, the nuclear spin splitting is negligible so states with different $m_I$ can be considered degenerate. This approximation means that two atoms in different hyperfine ground states will couple to Rydberg states at the same energy (both labeled as $\ket{r}$) and interact with each other via a van der Waals potential (Fig.~\hyperref[fig:sfig1]{\ref*{fig:sfig1}b}). 

In our quenches and lifetime measurements, we always start with a spin-polarized gas of either state $\ket{1}$ or $\ket{3}$ atoms (both states are essentially equivalent and we happen to have take some of our data in this paper using one or the other). However, for the interferometry measurements, we need to take into account the dressed interaction potential between two atoms in different hyperfine ground states which couple to $\ket{r}$.

\section{Interaction potential for two Rydberg dressed atoms in different ground states}\label{app:2dressing}
To obtain the dressed potential for two atoms in different ground states, we start by writing down the single-particle Hamiltonians for each atom in the $\{\ket{i},\ket{r}\}$ basis in the rotating frame (where $i\in\{1,2\}$ labels the ground states):
\begin{align}
    \hat{H}_1 = 
    \begin{pmatrix}
        0 & \Omega/2 \\
        \Omega/2 & -\Delta
    \end{pmatrix}
    \quad\mathrm{and}\quad
    \hat{H}_2 = 
    \begin{pmatrix}
        0 & \Omega/2 \\
        \Omega/2 & -(\Delta+\Delta_0)
    \end{pmatrix}
\end{align}
Using these and the interaction potential between two atoms in the Rydberg state separated by a distance $R$, $V(R) = -C_6/R^6$, we write down the two-particle dressing Hamiltonian as
\begin{align}
\hat{H}_{\textrm{dr}}(R) = \hat{H_1}\otimes\hat{\mathbb{I}} + \hat{\mathbb{I}}\otimes\hat{H}_2 + V(R)(\ket{r}\bra{r}\otimes\ket{r}\bra{r}).
\end{align}
We calculate the dressed potential by solving for the eigenenergy of the eigenstate with maximum overlap with the bare ground state $\ket{1,2}$. This can be done numerically, or using perturbation theory up to 4th order in $\Omega$ assuming $\Omega\ll\Delta$. In this limit, we find that the relevant eigenergy has the form
\begin{align}
    E(R) = &-\frac{\Omega^4(2\Delta+\Delta_0)}{16\Delta^2(\Delta+\Delta_0)^2}\left(\frac{1}{1 + \frac{(2\Delta+\Delta_0)R^6}{\abs{C_6}}}\right) \nonumber\\
    &+ \delta(\Omega, \Delta) + \delta(\Omega, \Delta + \Delta_0), \label{eq:eqB3}
\end{align}
where $\delta(\Omega,\Delta) = (-\Delta + \sqrt{\Delta^2+\Omega^2})/2$ is the single-particle light shift and the first term is the desired interaction potential.

\section{Many-body Ramsey interferometry}\label{app:ramsey}
We use the same many-body Ramsey interferometry technique as Ref.~\cite{Zeiher-Gross2016} to characterize the interaction potentials of the Rydberg dressed atoms. However, because the splitting between the ground states we use in the Ramsey interferometry is only $\SI{75.806(3)}{\MHz}$ and the detunings we use are between \SI{30}{\MHz} and \SI{100}{\MHz}, we need to take into account the dressing of both states. Since the experiments are performed in the frozen-gas regime, we rewrite the dressing Hamiltonian as a spin Hamiltonian. For our interferometer, we use hyperfine ground states $\ket{1}\equiv\ket{\uparrow}$ and $\ket{2}\equiv\ket{\downarrow}$. The many-body dressing Hamiltonian is
\begin{align}
    \hat{H}_{\textrm{dr}} =& \sum_i\left(\delta_i^{\uparrow}\hat{\sigma}_{\uparrow\uparrow}^{(i)} + \delta_i^{\downarrow}\hat{\sigma}_{\downarrow\downarrow}^{(i)}\right) \nonumber\\
    &+ \frac{1}{2}\sum_{i\neq j}\Big{(}V_{ij}^{\uparrow\uparrow}\hat{\sigma}_{\uparrow\uparrow}^{(i)}\hat{\sigma}_{\uparrow\uparrow}^{(j)} + V_{ij}^{\downarrow\downarrow}\hat{\sigma}_{\downarrow\downarrow}^{(i)}\hat{\sigma}_{\downarrow\downarrow}^{(j)} \nonumber\\
    &+ V_{ij}^{\uparrow\downarrow}\hat{\sigma}_{\uparrow\uparrow}^{(i)}\hat{\sigma}_{\downarrow\downarrow}^{(j)} + V_{ij}^{\downarrow\uparrow}\hat{\sigma}_{\downarrow\downarrow}^{(i)}\hat{\sigma}_{\uparrow\uparrow}^{(j)}\Big{)},
\end{align}
where $\delta_i^\alpha$ is the single-particle light shift for spin $\alpha$ at site $i$, $V_{ij}^{\alpha\beta}$ is the Rydberg dressed potential between spins $\alpha$ and $\beta$ at sites $i$ and $j$, and $V_{ij}^{\uparrow\downarrow} = V_{ij}^{\downarrow\uparrow}$. Using the relations $\hat{\sigma}_{\uparrow\uparrow}^{(i)} = \frac{1}{2}\left(\hat{\mathbb{I}}+\hat{\sigma}_z^{(i)}\right)$ and $\hat{\sigma}_{\downarrow\downarrow}^{(i)} = \frac{1}{2}\left(\hat{\mathbb{I}}-\hat{\sigma}_z^{(i)}\right)$, we can rewrite the Hamiltonian as
\begin{align}
    \hat{H}_{\textrm{dr}} &= H_0 \nonumber\\
    &\phantom{=} + \frac{1}{2}\sum_i \left(\delta_i^{\uparrow} -\delta_i^{\downarrow} +\frac{1}{2}\sum_{j\neq i}\left(V_{ij}^{\uparrow\uparrow} - V_{ij}^{\downarrow\downarrow} \right) \right)\hat{\sigma}_z^{(i)} \nonumber\\
    &\phantom{=} + \frac{1}{8}\sum_{i \neq j}\left(V_{ij}^{\uparrow\uparrow} + V_{ij}^{\downarrow\downarrow} - 2V_{ij}^{\uparrow\downarrow} \right)\hat{\sigma}_z^{(i)}\hat{\sigma}_z^{(j)} \\
   \hat{H}_{\textrm{dr}} &= H_0 + \frac{1}{2}\sum_i \delta_i^*\hat{\sigma}_z^{(i)} + \frac{1}{8}\sum_{i \neq j}V_{ij}^*\hat{\sigma}_z^{(i)}\hat{\sigma}_z^{(j)},
\end{align}
where $H_0$ is an energy offset, the second term is a longitudinal field of strength $\delta_i^*$ dominated by the single-particle light shifts, and the third term is an effective interaction term with strength $V_{ij}^*$. Similar to what is done in the supplement of Ref.~\cite{Zeiher-Gross2016}, we can calculate various observables for different pulse sequences in terms of the accumulated phases $\phi_i = \int_0^{\tau}\delta_i^*(t)\mathrm{dt}$ and $\Phi_{ij} = \int_0^{\tau}V_{ij}^*(t)\mathrm{dt}$ over the length $\tau$ of the dressing pulse.

For a $\pi/2-\tau-\pi/2$ pulse sequence, the observable is the expected single-component density $\hat{\sigma}_{\uparrow\uparrow}^i=\ket{\uparrow}\bra{\uparrow}$ which can be calculated to be:
\begin{align}
    \avg{\hat{\sigma}_{\uparrow\uparrow}^i} = \frac{1}{2} - \frac{1}{2}\cos(\phi_i)\prod_{j\neq i}\cos\left(\frac{\Phi_{ij}}{2}\right).
\end{align}

For a spin echo $\pi/2-\tau-\pi-\tau-\pi/2$ pulse sequence, the observable is the single-component density correlation which can be calculated to be:
\begin{align}
    \avg{\hat{\sigma}_{\uparrow\uparrow}^i\hat{\sigma}_{\uparrow\uparrow}^j}_C =& \frac{1}{8}\left(\prod_{k\neq i,j}\cos\Phi_{k,ij}^{(+)} +\prod_{k\neq i,j}\cos\Phi_{k,ij}^{(-)}\right)\nonumber\\
    &-\frac{1}{4}\cos\Phi_{ij}^2\prod_{k\neq i,j}\cos\Phi_{ik}\cos\Phi_{jk},
\end{align}
where $\Phi_{k,ij}^{(\pm)}=\Phi_{ik}\pm\Phi_{jk}$ and $\Phi_{ii} = 0$.

\section{Dependence of lifetime on atom number at fixed density}\label{app:lifetimeVnumber}
\begin{figure}
\includegraphics[width=\columnwidth]{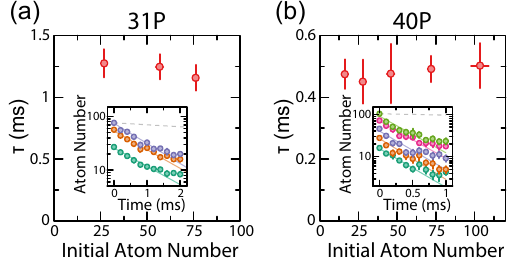}
\caption{{\bfseries Dependence of lifetime on atom number at fixed density.} (a) Initial lifetime for 2D systems with different initial atom number dressed to $31P$. Measurements are made in a 2D rectangular system of small width $\sim\SI{4}{a_{latt}}$ and variable length along the dressing beam direction. We observe no strong dependence on the atom number. The Rabi frequency is approximately constant over the entire system. For this data, $\Omega=2\pi\times\SI{7.02(5)}{\MHz},\Delta=2\pi\times\SI{60}{\MHz}$ and $n=0.8$. (b) Same as in (a) but for systems dressed to $40P$. For this data, $\Omega=2\pi\times\SI{5.6(2)}{\MHz},\Delta=2\pi\times\SI{40}{\MHz}$ and $n=0.55$. (insets) Raw data with exponential fits to extract the initial decay rate. Experimental error bars correspond to standard error of the mean. \label{fig:sfig2}}
\end{figure}

In our search for a suitable Rydberg state to use for our dressing experiments, we explored many different principal quantum numbers. We eventually chose $28P$ because it gave us a good ratio between the measured collective lifetime and the theoretical single-particle lifetime, while also having a large enough $C_6$ to achieve strong nearest-neighbor interactions in the lattice. We explored larger principal quantum numbers but found much shorter lifetimes than the expected values. One possible reason is the coupling to neighboring pair-potentials that have non-zero overlaps with the target state at close distances (Fig.~1 of main text). However, the general behavior of the many-body lifetimes with atom number and geometry of the cloud remained the same over significantly different principal quantum numbers. In particular, the lifetime showed no strong dependence on the atom number at fixed density over the range we could explore in the experiment. Fig.~\ref{fig:sfig2} shows the initial lifetime vs. the initial atom number for 2D systems $\SI{4}{a_{latt}}$ wide and of variable length along the direction parallel to the dressing beam for the $31P$ and $40P$ Rydberg states.

\section{Atom loss during charge density wave dynamics}\label{app:quenchlifetime}
\begin{figure}
\includegraphics[width=\columnwidth]{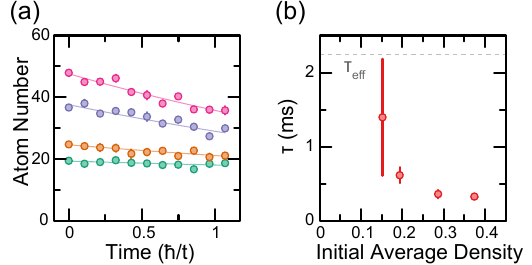}
\caption{{\bfseries Atom loss during charge density wave dynamics.} (a) Atom number decay over the quenches shown in Fig 5 of main text. Each color represents a set with a different initial density. Circles are measured data with error bars and lines are simple exponential decay fits. The dressing parameters were $\Omega = 2\pi\times \SI{6.99(8)}{\MHz}$ and $\Delta = 2\pi\times\SI{30}{\MHz}$. (b) Lifetime vs. the initial average density of the charge density wave as extracted from the data in (a). This behavior is in agreement with our observations shown in Fig.~3 of main text. Experimental error bars correspond to standard error of the mean. \label{fig:sfig3}}
\end{figure}

We observe an atom loss of $\sim30 \%$ for the longest evolution times for the dataset with the maximum initial density and interaction strength. For the dataset where interaction was varied by changing the dressing laser intensity, the lifetime gets longer for smaller interactions due to the reduction of the Rydberg dressing parameter $\beta=\frac{\Omega}{2\Delta}$. For the dataset where the initial density was varied at fixed interaction strength, the lifetime increased for lower initial densities (Fig.~\ref{fig:sfig3}). These measurements are in accordance with our observed density dependent lifetime measurements shown in Fig. 3 of main text.

\pagebreak



%

\end{document}